# Uncovering Causal Drivers of Energy Efficiency for Industrial Process in Foundry via Time-Series Causal Inference

Zhipeng Ma[1][0000-0002-4049-539X], Bo Nørregaard Jørgensen[1][0000-0001-5678-6602], Zheng Grace Ma[1][0000-0002-9134-1032]

[1] SDU Center for Energy Informatics, the Maersk Mc-Kinney Moller Institute, Faculty of Engineering, University of Southern Denmark, DK-5230 Odense, Denmark
`{zhma, bnj, zma}@mmmi.sdu.dk`

**Abstract.** Improving energy efficiency in industrial foundry processes is a critical challenge, as these operations are highly energy-intensive and marked by complex interdependencies among process variables. Correlation-based analyses often fail to distinguish true causal drivers from spurious associations, limiting their usefulness for decision-making. This paper applies a time-series causal inference framework to identify the operational factors that directly affect energy efficiency in induction furnace melting. Using production data from a Danish foundry, the study integrates time-series clustering to segment melting cycles into distinct operational modes with the PCMCI+ algorithm, a state-of-the-art causal discovery method, to uncover cause-effect relationships within each mode. Across clusters, robust causal relations among energy consumption, furnace temperature, and material weight define the core drivers of efficiency, while voltage consistently influences cooling water temperature with a delayed response. Cluster-specific differences further distinguish operational regimes: efficient clusters are characterized by stable causal structures, whereas inefficient ones exhibit reinforcing feedback loops and atypical dependencies. The contributions of this study are twofold. First, it introduces an integrated clustering-causal inference pipeline as a methodological innovation for analyzing energy-intensive processes. Second, it provides actionable insights that enable foundry operators to optimize performance, reduce energy consumption, and lower emissions.

## 1 Introduction

The iron and steel industry, a cornerstone of modern manufacturing, stands as one of the largest contributors to global greenhouse gas emissions, accounting for approximately 28.71% of industrial emissions in 2023 [1-2]. Despite efforts such as electrification and energy-efficient practices, the sector remains off-track to meet the International Energy Agency's (IEA) Net Zero Emissions by 2050 scenario [3]. Foundries, which supply essential metal castings for automobiles, infrastructure, and consumer goods, face the dual challenge of meeting rising demand while reducing their environmental footprint. In Europe, this challenge is amplified by strict sustainability regulations and high operational costs [4].

Crude steel production has doubled between 2000 and 2024 [5], driven by rapid industrial growth and infrastructure development. The melting process accounts for approximately 55% of the total energy consumption [6], making it a primary target for efficiency gains. However, systematic improvement remains difficult. Operations depend heavily on the tacit knowledge of furnace operators, while diverse melting patterns complicate performance comparison. The absence of standardized methods for analyzing or benchmarking melting practices further constrains progress.

Recent advances in digitalization have enabled the extensive collection of time series data from furnace operations [7]. While descriptive and predictive analytics can highlight correlations, they often fail to uncover the underlying drivers of efficiency. Without causal insights, interventions risk being ineffective or even counterproductive. Identifying causal relationships between operational variables and energy outcomes is therefore essential for guiding evidence-based strategies.

Causal inference has found growing applications in industrial systems. For instance, the study in [8] has proposed a root-cause analysis framework for complex processes, enabling timely corrective interventions. In [9], causal relationships among contributing factors and the main dimensions of accident consequences involving hazardous materials were examined to enhance safety learning from major incidents. More recently, the study in [10] has developed a causal network–based fault diagnosis framework that mitigates smearing effects while accurately identifying fault variables and their contribution rates, offering improved interpretability and scalability. However, few studies have investigated foundry production processes, and even fewer have applied causal analysis to energy efficiency. This gap highlights the need for systematic approaches that move beyond correlation to uncover the causal drivers of energy performance in foundry operations.

Building on this research gap, this paper aims to investigate the causal drivers of energy efficiency in foundry melting processes using time series causal inference. Real-world furnace operation data are analyzed to identify key operational variables and their interactions that directly influence energy consumption. In previous studies, clustering has been conducted to differentiate melting operations into distinct groups [11]. Building on this idea, the present work applies an enhanced PCMCI+ algorithm [12-13] within each cluster to uncover causal dependencies among process variables. The effectiveness of the method is demonstrated through a case study of a Danish foundry. The analysis reveals that causal relationships vary across operating conditions, with some factors, such as cooling water dynamics, furnace temperature, and the interaction between current and power.

The remainder of this paper is structured as follows. Section 2 introduces the background of foundry operations. Section 3 presents the methodology, including the time-series causal inference algorithm and data processing pipeline. Section 4 describes the case study and datasets, while Section 4.3 reports the experiments and results. Section 6 provides a discussion of the findings, and Section **Error! Reference source not found.** concludes the paper.

## 2     Background of Foundry Operations

The fundamental process flow of foundry production is illustrated in Fig. 1 [2, 6]. As shown, several sequential steps are required to produce the final casting workpieces.

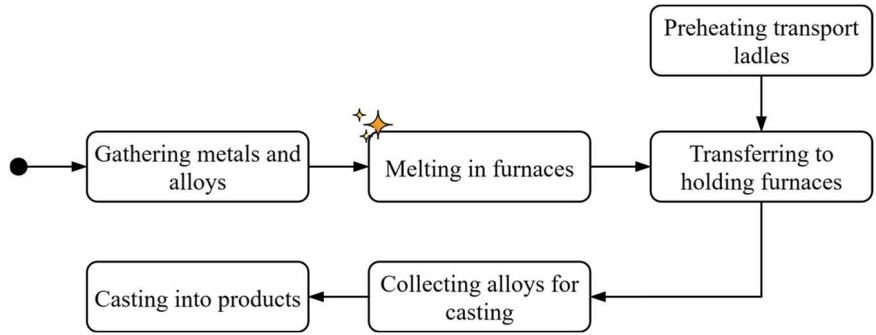

**Fig. 1.** Overview of the foundry production process [2, 6].

Production begins with the preparation of raw materials, consisting of iron and selected alloys, including the recycled scrap of varying quality grades. These inputs are then added into an induction furnace, where they are heated until the desired melting temperature is achieved, with adjustments made for ferromagnetic losses. Once molten and achieving the setting temperature, the metal is transferred into a pre-heated ladle. At this stage, doping may be performed to refine the composition based on the production recipe before the melt is transferred to the holding furnace for temporary storage, marking the completion of the melting stage.

The forming stage follows, beginning with the withdrawal of molten alloy from the holding furnace into another pre-heated ladle. Additional alloying doping may be applied here to meet customer-specific requirements. The prepared melt is then poured into sand molds through automated molding equipment, where it solidifies into the desired shapes. Finally, the castings are transferred to a cooling line, which completes the primary forming stage of the foundry process.

Foundry production is recognized as a highly energy-intensive activity, relying on substantial amounts of electricity, natural gas, and other fuels to melt and process metal alloys [14]. Previous research has examined the major sources of energy demand in such operations, highlighting stages such as melting, casting, heat treatment, and material handling. Among these, heat treatment and melting together have been reported to account for nearly 60% of the total energy use in aluminum foundries [15], while in gray iron foundries the melting stage alone contributes roughly 55% [16]. Given its dominant share of overall consumption, the melting process is selected as the focus of analysis in this study.

## 3   Methodology

This paper applies a time series causal inference algorithm to analyze energy efficiency patterns in melting processes. Section 3.1 presents the causal inference framework, including enhancements to the PCMCI+ algorithm, and Section 3.2 outlines the data processing pipeline developed for implementing the analysis.

### 3.1   Time-Series Causal Inference

Causal inference aims to identify cause-effect relationships between variables, moving beyond correlation-based analysis. In time series settings, the objective is to uncover how past values of one or more variables influence the future evolution of others. Unlike static data, time series present unique challenges due to autocorrelation, temporal dependencies, and potential confounding effects across multiple time scales [17-18].

In 1969, Clive Granger proposed the Granger causality test (GC) to determine whether a time series is useful for forecasting the other [19]. This was the first causal model for time-dependent systems. In recent years, with the progress of deep learning, statistics, and interdisciplinary research, many other state-of-the-art data-driven and theory-driven mechanisms have been developed [12-13, 20-22]. Among them, the PCMCI algorithm [12] is a power method that achieves high accuracy in real-world applications.

The PCMCI algorithm [12] quantifies the causal strength of large-scale nonlinear time-series datasets based on the CI tests, and its extension, PCMCI+ [13] includes the discovery of instantaneous (also called contemporaneous) relationships to improve the reliability of causal detection. Compared to generic PC/FCI-style approaches, PCMCI+ [13] optimizes conditioning sets and leverages momentary conditional independence (MCI) tests, which improves recall and controls false positives in the presence of strong autocorrelation, which is common in real-world time series. Therefore, this study adopts the PCMCI+ as the time-series causal discovery algorithm due to its superiority in complex nonlinear systems and its unique optimization in instantaneous links.

The PCMCI+ algorithm contains two main steps: condition selection with PC algorithm and MCI tests. In the first stage, the goal of the PC phase is to remove the conditions on irrelevant variables to reduce the dimensionality of conditioning sets for MCI tests. Considering the time series **X**, all time-lagged pairs $(X_{t-\tau}^i, X_t^j), \tau = 1, \dots, \tau_{\max}$ are tested given the conditions containing only the strongest $p$ links of $X_t^j$ in each $p$-iteration. The resulted adjacency set is denoted as $\hat{\mathcal{P}}(X_t^j)$. This processing removes the noise from the condition set to accelerate the following MCI step.

In the MCI step, the contemporaneous links $(X_t^i, X_t^j)$ for all $X_t^i \neq X_t^j$ are supplemented, and the orientations of the linkage are determined by three rules aimed at orienting collider triples, orienting chain triples, and avoiding cycles respectively. Next, the MCI tests are conducted given all lagged links from $\hat{\mathcal{P}}(X_t^j)$ and all contemporaneous links. Finally, the causal strength graphic is generated by removing all links that do not pass the MCI tests.

The causal strength is determined by the CI test. Each CI test provides a corresponding test statistic, which is interpreted as the causal strength. A larger absolute value of this statistic indicates a stronger causal relationship. Table 1 summarizes the CI tests implemented in the PCMCI+ algorithm [23], applicable to continuous variables characterized by diverse marginal distributions and noise structures, as frequently encountered in real-world datasets.

**Table 1.** Assumptions and applicability for different CI tests [23].

| No. | CI test | Assumptions and applicability |
|---|---|---|
| 1 | RobustParCorr | Suitable for univariate, continuous variables with linear relationships, robust to varying marginal distributions |
| 2 | ParCorrWLS | Suitable for univariate, continuous variables with linear relationships, can handle heteroskedasticity using weighted least squares |
| 3 | GPDC | Suitable for univariate, continuous variables with additive (including nonlinear) dependencies |
| 4 | CMIknn | Suitable for multivariate, continuous variables with more general dependencies via a permutation-based test |

The experiments in [24] demonstrate that, in complex systems, PCMCI+ combined with CMIknn test outperforms the other three CI tests in terms of both SHD and FDR, achieving an FDR close to zero. However, despite its low FDR, this method still fails to identify certain true causal links. In contrast, the other three CI tests, although they yield a higher number of false positives, are able to detect more causal relationships overall. To enhance the overall accuracy of PCMCI+, a hybrid approach integrating all four CI tests is designed.

This integration methodology is designed to enhance the accuracy of causal discovery by supplementing the results of PCMCI+ combined with the CMIknn test, while preserving its low FDR. The approach begins by computing four causal adjacency matrices using the four distinct CI tests listed in Table 1. These matrices are denoted as $\mathbf{W}_i$, where $i = 1,2,3,4$, corresponding to the order of CI tests in Table 1. The final integrated causal matrix $\mathbf{W}_{hybrid}$ is obtained through the following set-based operation:

$$\mathbf{W}_{hybrid} = (\mathbf{W}_1 \cap \mathbf{W}_2 \cap \mathbf{W}_3) \cup \mathbf{W}_4 \qquad (1)$$

This formulation retains high-confidence causal links agreed upon by the first three tests while incorporating additional links identified by the fourth test (CMIknn), which is known for its low FDR performance. $\mathbf{W}_{hybrid}$ is further refined by resolving bidirectional links. Specifically, if both $X^i \rightarrow X^j$ and $X^j \rightarrow X^i$ are present, the direction included in $\mathbf{W}_4$ is retained, and the other is removed. The result is a balanced integration that improves overall detection of true causal relationships without substantially increasing false positives.

### 3.2   A Data Processing Pipeline for Implementing Causal Inference

The overall data analysis pipeline adopted in this study is depicted in Fig. 2. It is structured into four stages. The first two stages, including operational segmentation and operational clustering, shown in dashed boxes, have been addressed in previous studies [11] and are adopted here as preparatory steps. Building on this foundation, the present work extends the analysis by applying time-series causal inference and post-hoc comparison to systematically uncover the causal drivers of energy efficiency in melting operations.

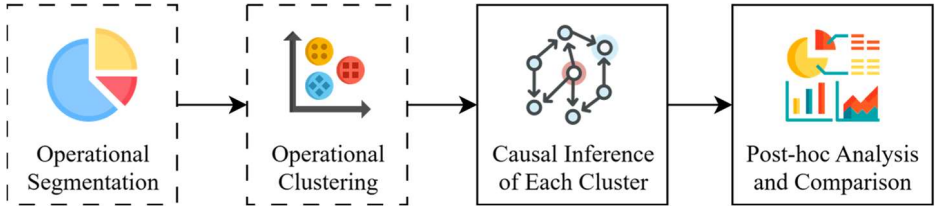

**Fig. 2.** Data analysis pipeline for identifying causal drivers of energy efficiency.

In the first stage, continuous furnace operation data are segmented into discrete melting operations. This step ensures that the analysis is based on well-defined production cycles rather than undifferentiated time series, thereby enabling consistency and comparability across operations.

The segmented operations are subsequently grouped into clusters according to similarities in their process characteristics. Clustering serves to differentiate heterogeneous operational modes that may arise from variations in metal compositions, melting duration, or energy intensity. By creating relatively homogeneous groups, this step reduces the influence of confounding factors and provides a clearer basis for causal inference.

Causal inference is then conducted separately within each cluster using an integrated PCMCI+ framework introduced in Section 3.1. This stage identifies direct causal relationships among process variables, accounting for temporal dependencies. Performing the analysis at the cluster level allows the detection of causal structures that are specific to distinct operational modes.

Finally, the causal graphs derived from individual clusters are subjected to post-hoc analysis and comparison. This synthesis highlights both common causal drivers and cluster-specific distinctions, offering a comprehensive view of how different operational patterns influence energy efficiency. The resulting insights provide a foundation for evidence-based strategies aimed at improving process sustainability and operational performance.

## 4   Case Study

This section presents the case study, beginning with background information in Section 4.1, followed by a description of the datasets in Section 4.2 and the subset selection in Section 4.3.

### 4.1 Case study overview

A large Danish foundry is selected as the case study for applying causal inference to analyze the factors influencing energy efficiency in process operations. This facility is one of the largest in Northern Europe, producing approximately 45,000 tonnes of casting products annually, which are exported to more than 25 countries. As part of its sustainability initiatives, the foundry has mapped its production-related energy consumption, revealing that 78.5% of the total annual energy use is electricity, with the majority consumed by melting and holding furnaces.

This case study provides a suitable context for examining causal relationships in melting operations, owing to the company's commitment to sustainable practices and its advanced induction furnace infrastructure. The facility employs induction furnaces equipped with monitoring sensors, enabling detailed analysis of operational patterns. The general production sequence in this case follows the process flow shown in Fig. 1. For the purposes of this study, only the melting stage is analyzed, as it represents the dominant contributor to electricity consumption. The melting stage involves six induction furnaces with comparable specifications, from which one representative furnace is selected as the focus of analysis in this paper.

### 4.2 Datasets Overview and Data Preparation

The dataset used in this study was collected from one of the six induction furnaces in the melting process over a continuous monitoring period from November 14, 2022 to April 30, 2024, with data recorded at 10-second intervals.

The dataset includes 10 variables as shown in Table 2.

**Table 2.** Variables and explanations of the collected dataset.

| No. | Variable | Unit | Explanation |
|---|---|---|---|
| 1 | Weight | kg | The weight of materials added to the furnace |
| 2 | State | bit | Array of bits, each indicating the current operational state of the furnace |
| 3 | Temperature | °C | Temperature of the melt in the furnace |
| 4 | Frequency | Hz | Operating frequency of the furnace |
| 5 | Voltage | Volt | Operating voltage of the furnace |
| 6 | Current | Ampere | Operating current of the furnace |
| 7 | Isolation resistance | kOmh | Electrical isolation resistance of the furnace |
| 8 | Energy act | kWh | Total energy consumed by each melting cycle |
| 9 | Energy specific | kWh/tonne | Specific energy consumption per tonne of the manufactured metal |
| 10 | Power | kW | Instantaneous electrical power usage of the furnace |
| 11 | Cooling water temperature | °C | Cooling water temperature, measured at ten different locations around the furnace |

| 12 | Cooling water quantity | L/min | Cooling water flow rate, measured at ten different locations around the furnace |

Each melting operation begins at ambient temperature and proceeds through a heating phase until the metal reaches the required melting point, followed by a heating phase until the metal reaches the required melting point. Once molten, the metal is tapped, after which the furnace cools in preparation for the next cycle. This process repeats continuously, with each operation cycle forming a distinct time series. Over the monitoring period, 3,927 melting operations were identified.

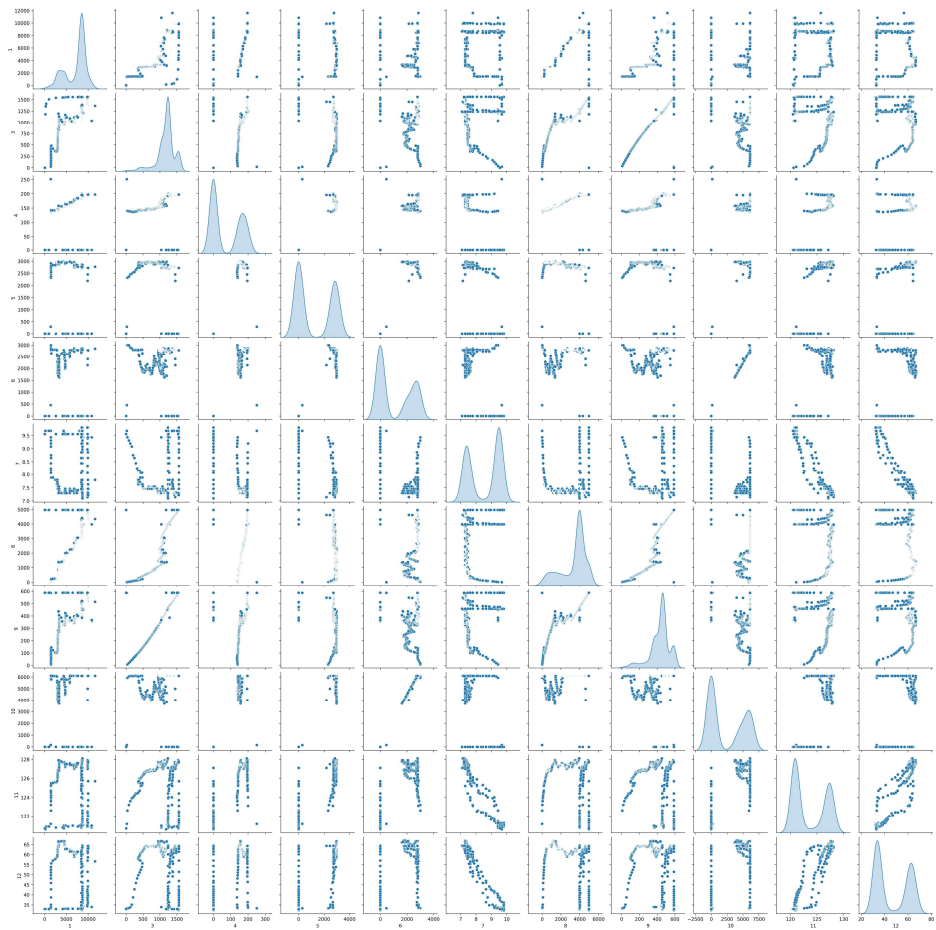

**Fig. 3.** Pairplot of data in a melting cycle.

To extract individual cycles, a segmentation pipeline has been developed and applied in previous research [11] based on domain-specific rules, combining temperature thresholds with time-based heuristics. Each sequence is treated as a self-contained melting cycle, beginning with preheating and ending after post-melt cooling. The resulting time series exhibit noise and occasional disturbances caused by operator interventions (e.g., mid-process alloying or furnace door opening), making the dataset highly representative of the variability and complexity of industrial practice. Fig. 3 illustrates an example of the data in a melting cycle using the pairplot. The variable "State" is removed because more than 99% of its values are missing [7].

Afterward, the temperature data is used for time-series clustering. The use of temperature as the clustering variable is motivated by its central role in the melting process. Temperature is directly coupled with electricity consumption, governs the thermodynamic and metallurgical transformations in the furnace, and thus provides a comprehensive representation of thermal and energy behavior during the cycle. Examining temperature trajectories enables the identification of distinct operational patterns and energy-consumption profiles, which can be associated with variations in charge composition, power input, furnace loading, and other process conditions.

The time-series clustering has been conducted in [11], and the melting cycles are clustered into seven groups. Table 3 summarizes key operational statistics for the seven clusters. The detailed algorithm development and result analysis of the clustering algorithm is published in [11].

**Table 3.** The statistical characteristics of each cluster [11].

| Attributes | Cluster 0 | Cluster 1 | Cluster 2 | Cluster 3 | Cluster 4 | Cluster 5 | Cluster 6 |
| --- | --- | --- | --- | --- | --- | --- | --- |
| Cardinality | 1562 | 488 | 466 | 440 | 353 | 313 | 305 |
| Average production time [s] | 4882.7 | 4340.5 | 4500.3 | 6043.2 | 4641.2 | 4963.7 | 7481.9 |
| Average weight [tonne] | 9.7 | 9.98 | 9.98 | 9.73 | 9.5 | 9.64 | 9.25 |
| Average electricity consumption [kWh] | 5068.4 | 5093.0 | 5260.1 | 5041.6 | 4853.6 | 5031.6 | 5054.5 |
| Average electricity consumption per unit [kWh/tonne] | 389.2 | 325.0 | 355.6 | 415.2 | 359.9 | 341.7 | 389.8 |

## 4.3 Subset Selection

Prior to causal inference, a data selection step is necessary because the full dataset is too large to be processed with available computational resources. While all variables and observations could, in principle, contribute to process understanding, applying causal inference directly to the complete dataset would be prohibitively time-consuming and memory-intensive. Selecting representative subsets ensures computational feasibility while preserving the system's essential dynamics.

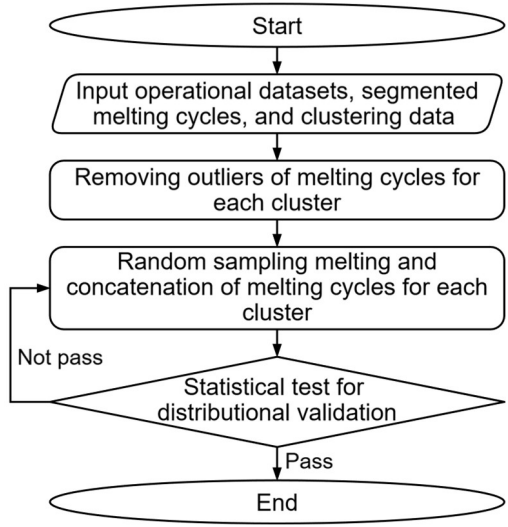

**Fig. 4.** Flowchart of data selection for time-series causal inference.

The procedure begins with outlier removal. Cycle-level variables such as production duration, total energy consumption, and specific energy consumption per tonne may contain extreme values caused by operational anomalies or sensor errors. Removing these values yields a dataset that reflects typical operational behavior and prevents atypical cycles from dominating the analysis.

Next, a random subset of melting cycles is sampled from each cluster. Sampling without replacement prevents duplication and preserves diversity, while randomness mitigates systematic bias. To ensure representativeness, the subset is validated against the original cluster distribution. Two complementary metrics are employed: Earth Mover's Distance (EMD) [25], which measures distributional shifts in individual features, and Maximum Mean Discrepancy (MMD) [26], which assesses multivariate distributional similarity in a reproducing kernel Hilbert space and is sensitive to higher-order and temporal differences. By enforcing thresholds on both metrics, the selected subsets remain statistically consistent with the original data.

The final output is a validated collection of subsets, one per cluster, concatenated into the dataset used for causal inference. This approach balances two objectives: maintaining the statistical fidelity of the original dataset and reducing its scale to enable tractable analysis. The resulting dataset provides a robust foundation for uncovering causal relationships in melting operations, ensuring both interpretability and reliability of the conclusions.

## 5     Results

This section presents the time series causal inference experiments and results for the seven identified clusters in Section 5.1, followed by the post-hoc analysis and cross-cluster comparison in Section 5.2.

## 5.1 Results of Causal Inference

In this study, the Integrated PCMCI+ algorithm is applied to each of the seven clusters. Given the substantial data volume in each cluster, ranging from approximately 500,000 to 1 million data points, analyzing the complete set of operations is computationally inefficient. To address this, about 5% of the operations is selected from each cluster and validated based on the pipeline in Section 4.3. The sampled operations were then integrated to form a representative sequence for causal analysis.

Each representative sequence is analyzed through the Integrated PCMCI+ algorithm. The results are illustrated in Fig. 5. Subfigures 4(a)–(g) depict the causal relationships among variables for cluster 0-6, respectively. Each node (black circle with a number) represents a variable listed in Table 2. Directed edges denote causal relationships (e.g., a → b indicates a causes b). The values along the edges represent the time lag with '1' denoting 10 seconds. The color of each edge reflects causal strength according to the color bar, while edge width also encodes strength, with wider edges indicating stronger relationships.

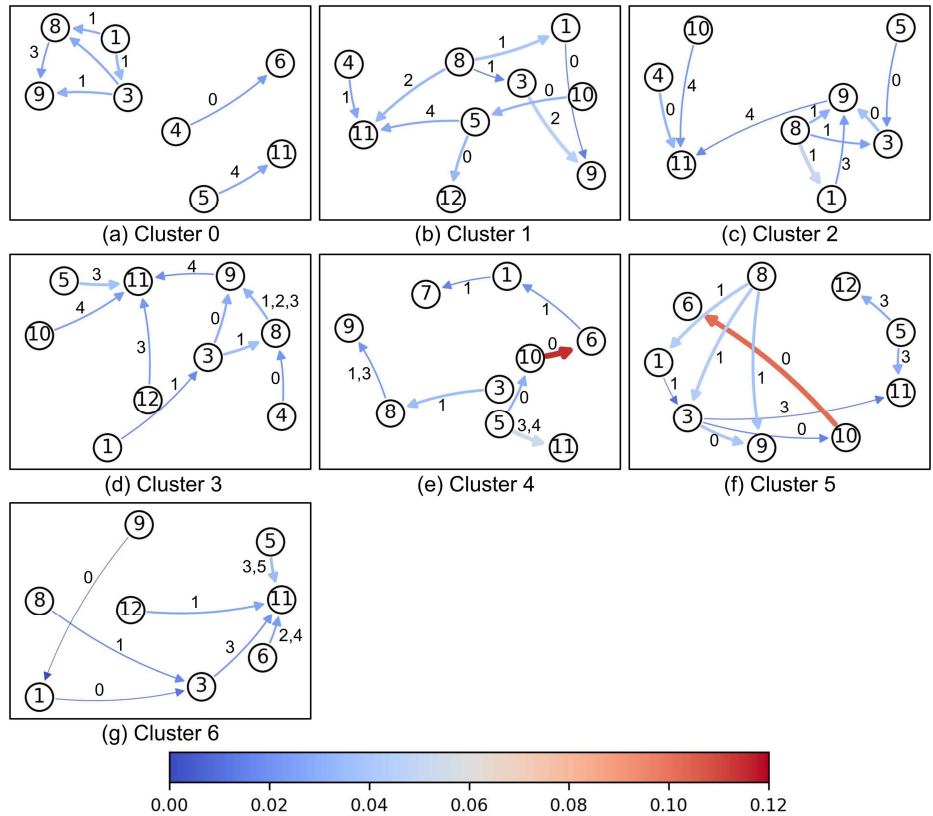

**Fig. 5.** Causal relationships within each cluster.

As illustrated in Fig. 5, the causal relationships differ across clusters. A more detailed analysis of these patterns is provided in Section 5.2.

**5.2   Post-hoc Analysis and Comparison**

This section provides a detailed analysis and interpretation of the causal relationships identified across clusters as illustrated in Fig. 5. The aim is to examine both the differences and commonalities in causal structures, thereby clarifying how efficiency drivers vary under distinct operational regimes.

Table 4 summarizes the occurrence frequency of each causal pair across the different clusters. The causal pair (a, b) represents that the variable a causes the variable b.

Table 4. Occurrence frequency of each causal pair across clusters.

| Occurrence Frequency | Causal Pairs |
| --- | --- |
| 6 | (5, 11) |
| 5 | (3, 9), (8, 9) |
| 4 | (1, 3), (8, 3) |
| 3 | (3, 8), (8, 1) |
| 2 | (1, 9), (3, 11), (4, 11), (5, 12), (9, 11), (10, 6), (10, 11), (12, 11) |
| 1 | (1, 7), (1, 8), (3, 10), (4, 6), (4, 8), (5, 3), (5, 10), (6, 1), (6, 11), (8, 11), (9, 1), (10, 5) |

The most frequent causal pair, *voltage → cooling water temperature* (5→11), appears in six clusters, confirming the consistent influence of electrical load on the cooling system. This relationship is characterized by a time lag of at least 30 seconds, indicating a delayed response of the cooling system to changes in electricity consumption. This link highlights the interplay between core furnace operations and auxiliary cooling systems, showing that higher electrical demand reliably translates into increased thermal stress on the cooling circuit.

Energy efficiency is primarily governed by two robust causal relations: *total energy consumption → specific energy consumption* (8→9) and *furnace temperature → specific energy consumption* (3→9), each recurring in five clusters. These links indicate that efficiency depends jointly on the amount of energy supplied and the thermal state of the melt. Two additional high-frequency relations, *weight → furnace temperature* (1→3) and *total energy consumption → furnace temperature* (8→3), occur in four clusters each. Notably, while four clusters support 8→3, three clusters display the reverse, 3→8, suggesting bidirectional interactions between temperature and total energy. This interplay among energy consumption, temperature dynamics, and material input defines a stable causal core that underpins furnace performance across operational regimes.

The causal graphs demonstrate that causal relationships differ across clusters. Cluster 0, which contains the largest number of melting cycles, exhibits the following causal structures. *Furnace temperature* (3) drives both *total energy consumption* (8) and *specific energy consumption* (9), underlining the role of thermal conditions in shaping energy use and efficiency. Interestingly, *weight* (1) causes *furnace temperature* (3), which likely reflects the alignment of temperature changes with material charging or melting progression in this regime.

Clusters 1, 4, and 5 are associated with higher efficiency. However, the DAGs for clusters 1 and 4 are unreliable, so emphasis is placed on cluster 5. In this cluster, *Furnace temperature* (3) causally affects *specific energy consumption* (9), *power* (10), and *cooling water temperature* (11), showing its broad influence on both electricity consumption and cooling systems. At the same time, *total energy consumption* (8) drives *weight* (1), *furnace temperature* (3), and *specific energy consumption* (9). The joint effect of *weight* (1) and *total energy consumption* (8) causing *furnace temperature* (3) suggests that material load and cumulative energy input together determine thermal operation state.

In contrast, cluster 6 represents the least efficient regime. Here, both *weight* (1) and *total energy* (8) causally determine *furnace temperature* (3), creating reinforcing dependencies. Additionally, *specific energy consumption* (9) causes *weight* (1), an atypical reversal that implies efficiency metrics might be entangled with material variables in non-productive ways. Such feedback loops indicate unstable or inefficient dynamics, where energy use is not effectively translated into productive melting.

## 6    Discussion and Conclusion

This study provides new insights into the cause-effect dynamics of melting operations, identifying which factors directly influence energy efficiency. By applying the integrated PCMCI+ algorithm, the analysis uncovers the directionality and interdependencies among operational variables, offering a more rigorous understanding than correlation-based methods.

A key finding is that cluster-specific causal networks reveal distinct efficiency drivers under different operational modes. Each of the seven melting clusters exhibits a unique causal structure, suggesting that furnaces employ diverse operational strategies. Equally important is the recurring role of cooling water dynamics: the link between flow rate and outlet temperature underscores the influence of auxiliary systems, highlighting opportunities for energy efficiency improvements beyond the furnace itself.

Methodologically, this work contributes to energy informatics by integrating causal discovery with clustering, providing a robust and explainable framework for analyzing heterogeneous industrial processes. Unlike prior studies that relied on correlations or expert judgment, the proposed approach infers directionality and interdependence directly from data, strengthening the analytical basis for industrial energy optimization.

Several limitations must be acknowledged. The findings are based on a single furnace at one foundry with a limited number of operations, which constrains generalizability. Differences in equipment, operational practices, and environmental conditions across sites may lead to alternative causal structures. Moreover, PCMCI+ identifies observational causal dependencies but cannot substitute for experimental validation. Future work should therefore extend the analysis to additional furnaces and longer datasets, and complement causal discovery with interventions or counterfactual simulations to validate and quantify the impact of identified drivers.

Despite these limitations, the framework offers practical value. By aligning interventions with cluster-specific causal structures, practitioners can implement targeted strategies rather than generic adjustments, improving both energy efficiency and process reliability. The approach also enables real-time classification of ongoing melts into known clusters, supporting context-specific decision-making. In doing so, it provides a generalizable template for causal analysis in complex, energy-intensive processes, with potential to reduce energy consumption, lower costs, and decrease greenhouse gas emissions.

**Acknowledgments.** This work is part of the project "Data-driven best-practice for energy-efficient operation of industrial processes - A system integration approach to reduce the CO2 emissions of industrial processes" (Case no.64020-2108) by the Energy Technology Development and Demonstration (EUDP) program, Denmark; part of the project titled "Dansk deltagelse i IEA IETS Task XIX Electrification in Industry Subtask 3", funded by EUDP (project number: 134251-549145); part of the project titled "Danish Participation in IEA IETS Task XVIII Digitalization, Artificial Intelligence and Related Technologies for Energy Efficiency and GHG Emissions Reduction in Industry Subtask 4", funded by EUDP (project number: 34251-549157); part of the project titled "Dansk deltagelse i IEA IETS Task XXIV – Procesintegration til Industriel Dekarbonisering", funded by EUDP (project number: 134251-549140).

**Disclosure of Interests.** The authors have no competing interests to declare that are relevant to the content of this article.

# References

1. International Energy Agency. World energy outlook 2024, https://www.iea.org/reports/world-energy-outlook-2024, last accessed 2025/08/24.
2. Howard, D.A., Værbak, M., et al.: Data-Driven Digital Twin for Foundry Production Process: Facilitating Best Practice Operations Investigation and Impact Analysis. In: Energy Informatics Academy Conference, pp. 259-73. Springer, (2024).
3. International Energy Agency. Global energy and climate model, https://www.iea.org/reports/global-energy-and-climate-model, last accessed 2025/08/24.
4. The European foundry industry. The European foundry industry 2023, https://eff-eu.org/wp-content/uploads/2024/12/CAEF-Co7_2023-complete.pdf, last accessed 2025/08/24.
5. Keramidas, K., Mima, S., Bidaud, A.: Opportunities and roadblocks in the decarbonisation of the global steel sector: A demand and production modelling approach. Energy and Climate Change 5, 100121 (2024).
6. Howard, D.A., Jørgensen, B.N., Ma, Z.: Identifying best practice melting patterns in induction furnaces: a data-driven approach using time series k-means clustering and multi-criteria decision making. In: Energy Informatics Academy Conference, pp. 271-88. Springer, (2023).
7. Ma, Z., Jørgensen, B.N., Ma, Z.G.: A systematic data characteristic understanding framework towards physical-sensor big data challenges. Journal of Big Data 11(1), 84 (2024).

8. Sun, Y., Qin, W., et al.: An adaptive fault detection and root-cause analysis scheme for complex industrial processes using moving window KPCA and information geometric causal inference. Journal of Intelligent Manufacturing 32(7), 2007-21 (2021).
9. Okki, S., Chebila, M., Nait-Said, R.: Causal Inference–Based Study of Key Contributors to Industrial Accidents. ASCE-ASME Journal of Risk and Uncertainty in Engineering Systems, Part A: Civil Engineering 10(1), 04023044 (2024).
10. Zhang, J., Rangaiah, G.P., et al.: A novel fault diagnosis framework for industrial production processes based on causal network inference. Industrial & Engineering Chemistry Research 63(21), 9471-88 (2024).
11. Ma, Z., Jørgensen, B.N., Ma, Z.G.: Discovering Operational Patterns Using Image-Based Convolutional Clustering and Composite Evaluation: A Case Study in Foundry Melting Processes. Information 16(9), 816 (2025).
12. Runge, J., Nowack, P., et al.: Detecting and quantifying causal associations in large nonlinear time series datasets. Science advances 5(11), eaau4996 (2019).
13. Runge, J.: Discovering contemporaneous and lagged causal relations in autocorrelated nonlinear time series datasets. In: Conference on uncertainty in artificial intelligence, pp. 1388-97. Pmlr, (2020).
14. Ma, Z., Friis, H.T.A., et al.: Energy Flexibility Potential of Industrial Processes in the Regulating Power Market. In: SMARTGREENS, pp. 109-15 (2017).
15. Nagapurkar, P., Paudel, S., Smith, J.D.: Improving Process Sustainability and Profitability for a Large US Gray Iron Foundry.
16. Salonitis, K., Jolly, M.R., et al.: Improvements in energy consumption and environmental impact by novel single shot melting process for casting. Journal of Cleaner Production 137, 1532-42 (2016).
17. Moraffah, R., Sheth, P., et al.: Causal inference for time series analysis: Problems, methods and evaluation. Knowledge and Information Systems 63(12), 3041-85 (2021).
18. Ma, Z., Kemmerling, M., et al.: A data-driven two-phase multi-split causal ensemble model for time series. Symmetry 15(5), 982 (2023).
19. Granger, C.W.: Investigating causal relations by econometric models and cross-spectral methods. Econometrica: journal of the Econometric Society, 424-38 (1969).
20. Runge, J., Gerhardus, A., et al.: Causal inference for time series. Nature Reviews Earth & Environment 4(7), 487-505 (2023).
21. Tank, A., Covert, I., et al.: Neural granger causality. IEEE Transactions on Pattern Analysis and Machine Intelligence 44(8), 4267-79 (2021).
22. Pamfil, R., Sriwattanaworachai, N., et al.: Dynotears: Structure learning from time-series data. In: International Conference on Artificial Intelligence and Statistics, pp. 1595-605. Pmlr, (2020).
23. https://github.com/jakobrunge/tigramite/, last accessed 2025/08/26.
24. Runge, J.: Conditional independence testing based on a nearest-neighbor estimator of conditional mutual information. In: International Conference on Artificial Intelligence and Statistics, pp. 938-47. PMLR, (2018).
25. Andoni, A., Indyk, P., Krauthgamer, R.: Earth mover distance over high-dimensional spaces. In: SODA, pp. 343-52 (2008).
26. Jia, X., Zhao, M., et al.: Assessment of data suitability for machine prognosis using maximum mean discrepancy. IEEE transactions on industrial electronics 65(7), 5872-81 (2017).